\documentstyle[preprint,aps]{revtex}
\begin{document}
\tighten
\title{Sub-threshold resonances in few--neutron systems}
\author{
S. A. Sofianos, S. A. Rakityansky\footnote{Permanent address:
Joint Institute  for Nuclear Research, Dubna, 141980, Russia},
G. P. Vermaak
}
\address{ Physics Department, University of South Africa,
 P.O.Box 392, Pretoria 0001, South Africa}
\maketitle
\begin{abstract}
Three-- and four--neutron systems are studied within the framework of
the hyperspherical approach with a local $S$--wave $nn$--potential. Possible
bound and resonant states of these systems are sought as zeros of three--
and four--body Jost functions in the complex momentum plane.
It is found that zeros closest to the origin correspond to sub--threshold
$(nnn)$ $\frac{1}{2}^-$ and $(nnnn)$  $0^+$  resonant states.
The positions of these zeros turned out to be sensitive to the choice
of the  $nn$--potential. For the Malfliet--Tjon potential they are
$E_0({}^3n)=-4.9-i6.9$\ (MeV) and $E_0({}^4n)=-2.6-i9.0$\ (MeV).
Movement of the zeros with an artificial increase of the potential strength
also shows an extreme sensitivity to the choice of potential. Thus, to
generate $^3n$ and $^4n$ bound states, the Yukawa potential needs to be
multiplied by 2.67 and 2.32 respectively, while for the Malfliet--Tjon
potential the required multiplicative factors are 4.04 and 3.59.
\end{abstract}
\vspace{.5cm}

\section{Introduction}

The question of the existence of pure neutron nuclei is a longstanding
problem, almost as old as nuclear physics itself. Many experiments have
been performed in the past in order to find di--, tri--, or tetra--neutron
bound states. On the theoretical front much effort has also been devoted
to understand the physics of few--neutron systems.\\

Early experimental and  theoretical investigations were controversial and
their findings varied from non-existence to `discovery' and `prediction'
of such states. Nowadays it is generally accepted that no few--neutron
nuclei exist. However, the
interest in pure neutron systems did not wane. Since  $(np)$, $(nnp)$,
$(npp)$, and $(nnpp)$ systems are bound, one could expect that due to
isotopic invariance a small increase in the neutron--neutron
attraction could generate bound states in  their isotopic partners
$(nn)$, $(nnn)$, and $(nnnn)$.  Indeed, this is the case for
the two--neutron system which requires only an additional 67\,keV
attraction for the formation of a bound di-neutron. For three and more
neutrons, however, the situation is more complicated due to the Pauli
principle. This becomes clear when one considers a multi--neutron
system in  even the simplest model, namely, that of neutrons moving
independently in a harmonic oscillator well. Since only two neutrons
can occupy the ground $S$--wave state of this well, the third, fourth,
and other neutrons must be in excited states and such a multi--neutron
system can form quasi--stationary states
instead of  bound ones. This encouraged research by both experimentalists
and theorists to locate resonances in  few--neutron systems \cite{NNDC}.\\

In spite of the abundance of experimental works that exploit different
reactions and methods, the situation concerning few--neutron resonances
remains unclear. In many cases certain irregularities in cross--sections
are observed, which can be attributed to such resonances, but their
positions and widths  have not been firmly established. As for
theoretical studies of few--body resonances, they are complicated
and cumbersome to the extend that these states have not been properly
investigated. Since bound states can be located more easily, it is
customary for theorists to increase the $nn$--potential until
a  bound multi--neutron system is supported.  It was found that
the interaction needs to be increased by a factor of between 2
and 4, depending on the  number of neutrons, the type of potential, and the
method used.\\

There are many different definitions of quantum resonances as well as many
methods of locating them \cite{orlov,kukulin,elander,con}. Though all
methods are physically justified and able to produce reasonable results,
only  few of them  can be considered as rigorous. These are the methods
based on the correspondence between resonances and singularities (poles)
of the $S$--matrix in the unphysical sheet of the complex energy plane.
Apart from being mathematically rigorous, such approaches enable one
to obtain the position and width of a resonance simultaneously with
the same accuracy, simply as coordinates of a pole on the energy
plane. And again, there are several different methods for locating the
$S$--matrix poles (a survey  of them can be found in Refs.
\cite{kukulin,elander}). However due to the existence of redundant
poles in the $S$--matrix it is preferable to search for  zeros of the
Jost function instead.\\

For two--body systems the Jost function is well--defined and an exact
method of calculating it has been  developed
\cite{nuovocim,zphys,nth9607028}. However, for many--body systems,
no such method exists and in fact  no rigorous definition of the Jost
function has  been given in a general form yet. Only for a special
class of many--body systems has a definition and method of calculating the
Jost function been proposed \cite{zphys}, namely, for  systems which cannot
form clusters of any type and thus only one allowed channel exists.
Some authors call them `democratic systems' \cite{gorb}.
Wave functions which describe such systems behave asymptotically
as diverging hyperradial spherical waves \cite{merkur}.
Clearly a multineutron system is one such system and  it  may
be investigated via the hyperspherical approach.\\

In this work we consider three-- and four--neutron systems in the minimal
approximation, $L=L_{min}$, of the hyperspherical harmonic approach with
local $S$--wave $nn$--potentials. Possible bound and resonant states
are sought as zeros of three--  and four--body Jost functions in the
complex plane of the momentum conjugate to the hyperradius. These
zeros are related to the total energy in a straightforward way.
The many--body Jost function for the hyperradial equation is defined
similarly  to the two--body case and is calculated by a method
which combines the variable constant and complex rotation methods
\cite{nuovocim}. The  zeros found represent the following
sub--threshold resonant states: $^3n(\frac{1}{2}^-)$,
$^4n(0^+)$, $^4n(1^+)$, and $^4n(2^+)$.\\

Our paper is organized as follows: In Sec. II we discuss the many--body
Jost function and in Sec. III we  give the required matrix elements of
the potential. In Sec. IV the complex coordinate rotation method
is used to achieve an analytic continuation of the Jost function
into the lower half of the momentum plane, while in Sec. V we describe
the physical input used in the calculations
and present our  results together with some discussions.
%
\section{Many--body Jost function}
Let $\{{\bf r}\}$ and $\{{\bf p}\}$ be the complete sets of Jacobi vectors
defining a multi--neutron configuration in coordinate and momentum space
respectively. The wave function $\Psi^a({\bf r},{\bf p})$ of an $N$--neutron
system can be expanded in terms of hyperspherical harmonics
$Y_{[L]}^a(\Omega)$ with unknown coefficient--functions
$u_{[L]}^a(r,{\bf p})$
\begin{equation}
\label{hex}
	\Psi^a({\bf r},{\bf p})=r^{2-3N/2}\sum_{[L]=[L_{min}]}^\infty
	Y_{[L]}^a(\Omega)u_{[L]}^a(r,{\bf p})\ ,
\end{equation}
where $\{a\}$ is the set of all conserved quantum numbers (total angular
momentum, isospin, and parity) and the multi-index $[L]$ represents the `grand
orbital'  $L$ and other nonconserved quantum numbers;  $r$ and $\Omega$,
$\{r, \Omega\}\equiv \{{\bf r}\}$, are  the  hyperradius and
hyperangles.  For two neutrons,
$N=2$,  the expansion is the usual partial--wave decomposition where
$[L]\equiv [\ell,m]$ is the angular momentum and its third component,
and $\ell_{min}=0$. When $N>2$, $L$ starts from a generally nonzero
value $L_{min}$ defined by the symmetry properties of the state under
consideration \cite{mishel}.\\

Substituting the expansion (\ref{hex}) into the Schr\"odinger
equation, one arrives at the following infinite system of coupled
hyperradial equations
\begin{equation}
\label{req}
	\left[\partial_r^2+p^2-\lambda(\lambda+1)/r^2\right]
	u_{[L]}^a(r,{\bf p})=\sum_{[L']}W_{[L][L']}^a(r)\, u_{[L']}^a(r,{\bf p})\ .
\end{equation}
Here
$
	[L]=[L_{min}], [L_{min}+1], \cdots
$
and  $p$ may be called `hypermomentum' since it is related
to the total energy $E$ and the neutron mass $m$ in the same manner as in
the two--body case, $p^2=2mE$ (we use units such that $\hbar=1$);
$\lambda$ is an analog of the orbital angular momentum,
\begin{equation}
\label{lambda}
	\lambda\equiv L+\frac32(N-2)\ ,
\end{equation}
which assumes half-integer values for uneven particle number $N$.
The potential matrix $W$ has
elements
\begin{equation}
\label{vmat}
	W_{[L][L']}^a(r)\equiv 2m\langle Y_{[L]}^a|
	\sum_{i<j}V_{ij}({\bf r})|Y_{[L']}^a\rangle\ ,
\end{equation}
where the integrals are over the hyperangles and $V_{ij}$ are the
two-body potentials.\\

The required boundary conditions for the differential equations
(\ref{req}) are, firstly, the solutions must be regular at the origin,
\begin{equation}
\label{orig}
	u_{[L]}^a(r,{\bf p})\to 0\ ,\qquad r\to 0\ ,
\end{equation}
and, secondly, at infinity  must be of some  physically motivated
form
\begin{equation}
\label{infty}
	u_{[L]}^a(r,{\bf p})\to U_{[L]}^a(r,{\bf p})\ ,\qquad r\to \infty\ .
\end{equation}
In the general case the boundary function $U_{[L]}^a(r,{\bf p})$ consists
of terms describing each  open channel. Since a multi--neutron system is a
`democratic' one, we shall construct $U_{[L]}^a$ as an one--channel
boundary function.\\

Instead of matching $u_{[L]}^a(r,{\bf p})$ to
$U_{[L]}^a(r,{\bf p})$, we can secure the one--channel boundary condition
automatically, and in an exact form, by the following procedure. Let us
consider a general regular solution of  (\ref{req}), defined only
by the condition (\ref{orig}) and without any restrictions at large $r$.
Certainly many such solutions exist, each having different
behaviour at large distances. By choosing only those which are
linearly independent, we have a regular basis.
Any other solution which is regular and has a specific behaviour at
large $r$ must be a linear combination of this basis set. Thus, we
can find first the regular basis and then construct the required
physical solution.\\

Like any other basis, the regular basis may be chosen in an infinite
number of ways. Exploiting this freedom, we can choose the basis
regular solutions in a way which is most suitable for the subsequent
construction of a physical solution. Considering the system (\ref{req})
(which is always truncated at some $[L_{max}]$) as a matrix equation,
we see that it has as many independent regular column--solutions as the
column dimension (number of equations in the system). We may combine all
these linear independent columns in a square matrix $\|\phi_{[L][L']}^a\|$
of the form (for simplicity we drop the superscript $a$ )
\begin{equation}
\label{ansatz}
	\phi_{[L][L']}(r,p)=\frac12\left\{
	h_\lambda^{(+)}(pr)F_{[L][L']}^{(+)}(r,p)+
	h_\lambda^{(-)}(pr)F_{[L][L']}^{(-)}(r,p)\right\}\ ,
\end{equation}
where $h_\lambda^{(\pm)}$ are the Riccati--Hankel functions \cite{abram},
and $F_{[L][L']}^{(\pm)}(r,p)$ are new unknown functions. Since instead of
one unknown matrix $\phi_{[L][L']}$, we have introduced two matrices
$F^{(\pm)}_{[L][L']}$, we require an additional constraint. The most
convenient is the Lagrange condition
\begin{equation}
\label{lagrange}
	h_\lambda^{(+)}(pr)\partial_rF_{[L][L']}^{(+)}(r,p)+
	h_\lambda^{(-)}(pr)\partial_rF_{[L][L']}^{(-)}(r,p)=0\ ,
\end{equation}
which is standard in the variable--constant method for solving
differential equations \cite{mathews}.\\

Substituting the ansatz (\ref{ansatz}) into Eq. (\ref{req}) and using
the condition (\ref{lagrange}), we derive the following coupled
differential matrix equations of first order:
\begin{eqnarray}
\nonumber
	\partial_rF_{[L][L']}^{(+)}(r,p)=\phantom{+}
	\frac{h_\lambda^{(-)}(pr)}{2ip}
	 \sum_{[L'']}W_{[L][L'']}(r)\left\{
	h_{\lambda''}^{(+)}(pr)F_{[L''][L']}^{(+)}(r,p)+
	h_{\lambda''}^{(-)}(pr)F_{[L''][L']}^{(-)}(r,p)\right\},\\
\label{fpmeq}
\phantom{-}\\
\nonumber
	\partial_rF_{[L][L']}^{(-)}(r,p)=-
	\frac{h_\lambda^{(+)}(pr)}{2ip}
	\sum_{[L'']}W_{[L][L'']}(r)\left\{
	h_{\lambda''}^{(+)}(pr)F_{[L''][L']}^{(+)}(r,p)+
	h_{\lambda''}^{(-)}(pr)F_{[L''][L']}^{(-)}(r,p)\right\}.
\end{eqnarray}
These equations must be supplemented with appropriate boundary
conditions at $r=0$. In Ref. \cite{palumbo} it was shown
that for an arbitrary $N$--body system the fundamental system of
regular solutions of Eq. (\ref{req}) vanishes near  $r=0$
in such a way that
\begin{equation}
\label{palum}
	\lim_{r\to 0}\frac{\phi_{[L][L']}(r,p)}
		{r^{\lambda'+1}}=\delta_{[L][L']}\ .
\end{equation}
Thus, we can define the regular basis by the following boundary
condition
\begin{equation}
\label{bcond}
	\lim_{r\to 0}\frac{\phi_{[L][L']}(r,p)}{j_{\lambda'}(pr)}
	= \delta_{[L][L']}\ ,
\end{equation}
where $j_\lambda$ is the Riccati--Bessel function \cite{abram}. This is
in accordance with (\ref{palum}) and is a natural generalization of
the corresponding boundary condition of the two--body problem.\\

Since, by  definition, $\phi_{[L][L']}$ is regular at $r=0$, the
behaviour of the functions $F_{[L][L']}^{(+)}$ and $F_{[L][L']}^{(-)}$
of Eq. (\ref{ansatz}) near the origin is such that the singularities of
$h_\lambda^{(+)}(pr)$ and $h_\lambda^{(-)}(pr)$ compensate each other.
This can be achieved if $F^{(+)}$ and $F^{(-)}$ are identical as $r\to 0$, i.e.,
$$
     F_{[L][L']}^{(\pm)}(r,p)\mathop{\sim}\limits_{r\to0}
     F_{[L][L']}(r,p)\ ,
$$
for then
\begin{equation}
\label{e}
      \phi_{[L][L']}(r,p)\mathop{\sim}\limits_{r\to0}
      j_\lambda(pr)\ F_{[L][L']}(r,p)\ ,
\end{equation}
so that the boundary conditions, Eq. (\ref{bcond}),  become
\begin{equation}
\label{d}
     \lim\limits_{r\to0}\left[
	\frac{j_\lambda(pr)  F_{[L][L']}^{(\pm)}(r,p)}
	{j_{\lambda'}(pr) }  \right] =\delta_{[L][L']}\ .
\end{equation}
However, since $r=0$ is a singular point, for practical calculations
one needs to solve the system (\ref{fpmeq}) analytically on a small
interval $(0,\delta]$ and then  impose the boundary conditions at
$r=\delta$. Such an analytical solution can be easily found
by choosing $\delta$ to be small enough so that for $r\in(0,\delta]$
we may write
$$
	\partial_rF_{[L][L']}^{(\pm)}(r,p)=\pm\frac{1}{ip}
	h_\lambda^{(\mp)}(pr)\ W_{[L][L']}(r) \ j_{\lambda'}(pr)\ .
$$
For small $r$ the Riccati-Neumann function $n_\lambda$ is dominant in
$h_\lambda^{(\pm)}\equiv j_\lambda\pm in_\lambda$ and  thus
$$
	\partial_rF_{[L][L']}^{(\pm)}(r,p) \approx -\frac{1}{p}
	n_\lambda(pr)\ W_{[L][L']}(r)\ j_{\lambda'} (pr)\ .
$$
Upon integrating this (approximate) equation we  find
$$
	F_{[L][L']}^{(\pm)}(r,p) \approx -\frac{1}{p}\int
	n_\lambda(pr)\ W_{[L][L']}(r)\ j_{\lambda'} (pr)\ {\rm d}r
	+{\rm const}\ ,
$$
and if the arbitrary constant of integration is taken to be
$\delta_{[L][L']}$, we obtain for the short range behavior,
\begin{equation}
\label{f}
	F_{[L][L']}^{(\pm)}(r,p)\mathop{\approx}\limits_{r\to0}
	\delta_{[L][L']}-\frac{1}{p}\int n_\lambda(pr)
	\ W_{[L][L']}(r)\ j_{\lambda'}(pr){\rm d}r\ ,
\end{equation}
which obeys the condition (\ref{d}). In practical calculations the
last indefinite integral can be found analytically by using the leading
terms of series expansions of $n_\lambda,\,W_{[L][L']}$ and
$j_{\lambda'}$. In this way one finds that the second term of
Eq.~(\ref{f}) is regular at $r=0$  for $\lambda\le\lambda'$
(right upper corner and the diagonal of the  matrix) and may be singular
for $\lambda>\lambda'$ (left lower corner of the matrix). However,
this singularity is always compensated  by the presence of
$j_\lambda$ in Eq.~(\ref{e}). Thus the coupled equations,
Eqs.~(\ref{fpmeq}), along with the boundary conditions, Eqs.~(\ref{d})
and (\ref{f}), form a well-defined differential problem.\\

The regular basis obtained in the form of Eq. (\ref{ansatz}) is ideally
suited for constructing physical solutions for  one--channel problems.
Indeed, since the right hand sides of Eqs.~(\ref{fpmeq}) vanish together
with the potential, so do the derivatives $\partial_rF^{(\pm)}$, which in
turn implies that beyond some $r_{max}$ both function
$F_{[L][L']}^{(\pm)}(r,p)$ become practically constant and the
asymptotic behavior of $\phi$ is totally determined by the Riccati--Hankel
functions. On the other hand, in  one--channel problems we can have only
three types of physical solutions describing bound, Siegert, and
scattering  states. For all of them the boundary function
$U$ of Eq. (\ref{infty}) must be constructed from the Riccati--Hankel
functions which depend on the hyperradius and at large $r$ behave as
\cite{abram}
\begin{equation}
\label{hinf}
	h_\lambda^{(\pm)}(pr)\mathop{\longrightarrow}\limits
	_{r\to\infty}\mp i\exp[\pm i(pr-\frac{\lambda\pi}{2})]\ .
\end{equation}
Thus the physical solution is a linear combination of the form
$$
	u_{[L]}(r,{\bf p})=\sum_{[L']}\phi_{[L][L']}(r,p)A_{[L']}({\bf p})\ ,
$$
which at $r=r_{max}$ smoothly matches, and for $r>r_{max}$ automatically
coincides with, the boundary function $U$,
$$
	\frac12\left\{
	h_\lambda^{(+)}(pr)\sum_{[L']}F_{[L][L']}^{(+)}
	(r_{max},p)A_{[L']}({\bf p})+
	h_\lambda^{(-)}(pr)\sum_{[L']}F_{[L][L']}^{(-)}
	(r_{max},p)A_{[L']}({\bf p})
	\right\}=U_{[L]}(r,{\bf p})\ ,
$$
provided the correct coefficients $A_{[L]}({\bf p})$ are found.\\

In the present work we are concerned with bound and Siegert states. For both
of them each element of the column $u_{[L]}$ at large $r$ must be proportional
to $h_\lambda^{(+)}(pr)$ which exponentially decays when $p$ is on the positive
imaginary axis (bound state), or represents pure outgoing waves when $p$ is
in the fourth quadrant of the complex $p$--plane (resonant  Siegert
state). This can be achieved provided that
\begin{equation}
\label{FA}
	\sum_{[L']}F_{[L][L']}^{(-)}(r_{max},p)A_{[L']}(p)=0\ .
\end{equation}
Here the argument of $A$ is $p$ because the wave function of a bound
or Siegert state does not depend on the orientation of the incident
momenta $\{{\bf p}\}$.\\

The homogeneous system of equations (\ref{FA}) has a nontrivial solution if
and only if
\begin{equation}
\label{detF}
	{\rm det}\,\|F_{[L][L']}^{(-)}(r_{max},p)\|=0\ .
\end{equation}
The discrete points $p=p_{01},\,p_{02},\,\dots$  at which Eq.
(\ref{detF}) is fulfilled, are the spectral points corresponding to bound
and resonant states.\\

When the number of particles $N=2$, the functions $F^{(\pm)}(r,p)$ are
closely related to the Jost solutions and the limit of $F^{(-)}$,
\begin{equation}
\label{fF}
	f(p)=\lim_{r\to\infty}F^{(-)}(r,p)\ ,
\end{equation}
is the Jost function. By analogy, we can call
$\|F_{[L][L']}^{(-)}(\infty,p)\|$ the Jost matrix for a one--channel
$N$--body problem. In practical calculations instead of $r=\infty$ we can
always use $r=r_{max}$.
%
 \section{Matrix elements of the potential}
%
>From the simplest shell--model it follows that the lowest configurations
for three and four neutrons are $(0s)^2(0p)$ and $(0s)^2(0p)^2$ and so we
expect to find one and two neutrons with
$\ell=1$ respectively. More elaborate few--body analyses
\cite{gorb,badalyan,birger,glockle,offermann,bevela} corroborate this
simplistic argument and concluded that if the nuclei $^3n$ and $^4n$
were to exist, the most favorable quantum numbers $(J^\pi)$ would
be  $\frac12^-$ for $^3n$ and $0^+,\,1^+$, or $2^+$ for $^4n$.
This means that even if $^3n$ and $^4n$ do not exist, the resonant poles
corresponding to these states must be the closest to the origin of the complex
energy plane. In the present work we shall search only for these states.\\

Since in the $nnn$  and $nnnn$ systems one and two particles respectively
 have $\ell=1$, the minimal value of the grand orbital number for them
is $L_{min}=1$ for $nnn$ and $L_{min}=2$ for $nnnn$. The general rule
defining $L_{min}$ can be found in Ref. \cite{mishel}. It has been
pointed out in many papers
(see, for example, Refs. \cite{gorb,badalyan,simonov}) that the minimal
approximation $L=L_{min}$ where only the first equation of the
system (\ref{req}) is retained, provides an adequate description of the
properties of $^3$He and $^4$He nuclei and that the corresponding minimal
components, $u_{[L_{min}]}(r,p)$, of the wave functions contribute
$\sim95$\% to the total normalization integral. Therefore we employ
this approximation to investigate the analytical properties of
the multi--neutron Jost function and thus to shed some light on the
question of existence of  resonances in many--neutron systems.

The general form of the matrix elements $W_{[L_{min}][L_{min}]}$
for three and four neutron systems were given in Refs. \cite{birger} and
\cite{badalyan}. They are
\begin{eqnarray}
\label{v3m}
	W_{[L_{min}][L_{min}]}^{(nnn,1/2^-)}(r) &=&
	\frac{48}{\pi}\int_0^{\pi/2}\,d\theta\,\sin^4\theta\cos^2\theta
	\,V_{nn}(\sqrt{2}r\cos\theta)\ ,\\
\nonumber
\phantom{-}\\
\label{v0p}
	W_{[L_{min}][L_{min}]}^{(nnnn,0^+)}(r) &=&
	\frac{105\cdot33}{8\cdot16}\int_0^{\pi/2}\,d\theta\,
	\sin^5\theta\cos^2\theta
	\,(4\cos^4\theta-4\cos^2\theta\sin^2\theta \\
\nonumber
	 &&+\frac{13}{4}\sin^4\theta)V_{nn}(\sqrt{2}r\cos\theta)\ ,\\
\nonumber
\phantom{-}\\
\label{v1p}
    W_{[L_{min}][L_{min}]}^{(nnnn,1^+)}(r) &=&
	\frac{105\cdot33}{64}\int_0^{\pi/2}\,d\theta\,\sin^9
	\theta\cos^2\theta \,V_{nn}(\sqrt{2}r\cos\theta)\ ,\\
\nonumber
\phantom{-}\\
\label{v2p}
	W_{[L_{min}][L_{min}]}^{(nnnn,2^+)}(r) &=&
	\frac{105\cdot33}{80}\int_0^{\pi/2}\,d\theta\,
	\sin^5\theta\cos^2\theta
	\,(\cos^4\theta+\frac{25}{16}\sin^4\theta)V_{nn}
	(\sqrt{2}r\cos\theta)\ .
\end{eqnarray}
These potentials are employed in our search for sub--threshold
resonances.

\section{Complex rotation}
%
Within the minimal approximation, only the first of Eqs. (\ref{req})
remains and consequently we have only one pair of Eqs. (\ref{fpmeq}).
Therefore the problem is similar to the two--body one with
potentials (\ref{v3m}--\ref{v2p}) and  angular momenta obtained by Eq.
(\ref{lambda}), that is $\lambda=5/2$ for $(nnn)$ and $\lambda=5$ for
$(nnnn)$. The corresponding Jost function becomes an effective
two--body one.\\

It is known that starting with a two--body radial
Schr\"odinger equation in its ordinary form, the Jost function for a
long--range potential can be defined for ${\rm Im}\, \{p\}\ge 0$ only.
The potentials (\ref{v3m}--\ref{v2p}) are clearly of the long--range
type. This is a result of a  general rule in which  two--body potentials
when  sandwiched  between hyperspherical harmonics, aquire slowly decaying
tails (see Ref. \cite{mishel}). For example, even if we take $V_{nn}$ in
Eq. (\ref{v3m}) in the form of a square well, the resulting function
$W_{[L_{min}][L_{min}]}(r)$ at large $r$ behaves as $r^{-3}$
\cite{birger}. Thus, the limit (\ref{fF}) does not exist in
the fourth quadrant of the complex $p$--plane where we shall
search for possible resonances.\\

To overcome this difficulty we employ the complex rotation
method in the form developed in Refs. \cite{nuovocim,nth9607028}.
In Eqs. (\ref{fpmeq}) we replace the real hyperradius with a complex one,
viz.
$$
	r=x\exp(i\theta)\ ,\qquad x\ge0\ ,\qquad 0\le\theta<\frac{\pi}{2}\ ,
$$
\begin{eqnarray}
\nonumber
	\partial_x\,F^{(+)}(x,\theta,p) &=& \frac{e^{i\theta}}
	{2ip}h_\lambda^{(-)}(pxe^{i\theta})\ W(xe^{i\theta})\\
\nonumber
	&&\times \left[h_\lambda^{(+)}(pxe^{i\theta})\ F^{(+)}
	(x,\theta,p) + h_\lambda^{(-)}(pxe^{i\theta})
	\ F^{(-)}(x,\theta,p)\right]\ ,\\
\label{dfmin}
&&\phantom{------------------}\\
\nonumber
	\partial_x\,F^{(-)}(x,\theta,p) &=&-\frac{e^{i\theta}}
	{2ip}h_\lambda^{(+)}(pxe^{i\theta})\ W(xe^{i\theta})\\
\nonumber
	&&\times \left[h_\lambda^{(+)}(pxe^{i\theta})\ F^{(+)}(x,\theta,p)
	+ h_\lambda^{(-)}(pxe^{i\theta})\ F^{(-)}(x,\theta,p)\right]\ .
\end{eqnarray}
%
Such a rotation does not change the Jost function which is
$r$--independent. Meanwhile, it changes the functions $F^{(\pm)}$
to the effect that $F^{(-)}(r,p)$ can be defined above the line
$(-\infty e^{-i\theta},+\infty e^{-i\theta})$ in the complex $p$--plane.
Moreover, in Ref. \cite{nuovocim} it was shown that at all points
above this line the limit (\ref{fF}) exists and gives the correct Jost
function. Therefore, using a rotation with large enough $\theta$, we can
calculate the Jost function at the points of interest in the fourth
quadrant of the $p$--plane and thus the multineutron resonances, if any,
can be located.
%
\section{Results and discussion}
%
A multitude of $nn$--interactions can  be found in the literature. We
may  divide them into two classes, namely, those with and those without a
repulsion at small distances. In this work we performed calculations
using both types of potentials.\\

As a purly atractive potential we choose  the Yukawa type
$$
   V_{nn}(r)=-V_0\frac{e^{-\alpha_0r}}{r} \ ,
$$
with $V_0=54.7477\ {\rm MeV\ fm}$ and $\alpha_0=0.84034\ {\rm fm}^{-1}$.
>From the second class we employ the Malfliet-Tjon singlet
potential \cite{MTI-III}
$$
  V_{nn}(r)=V_1\frac{e^{-\alpha_1r}}{r}-V_2\frac{e^{-\alpha_2r}}{r} \ ,
$$
with $V_1=1438.72\ {\rm MeV\ fm}$, $V_2=513.968\ {\rm MeV\ fm}$,
$\alpha_1=3.11\ {\rm fm}^{-1}$, and $\alpha_2=1.55\ {\rm fm}^{-1}$.
Both these potentials describe a nucleon--nucleon interaction in the
$^1S_0$ channel and support a virtual di-neutron state. We note that we
have slightly adjusted the depth of the Yukawa potential so that the
virtual bound state it generates is the same as for the Malfliet--Tjon
potential.  This virtual  bound state energy is given in Table 1.\\

Solving Eqs. (\ref{dfmin}) numerically with the potential matrices
(\ref{v3m}-\ref{v2p}), we found that, beyond $x_{max}=40$~fm, the function
$F^{(-)}(x,\theta,p)$ becomes practically constant (varying only in the
8-th digit). Further, the value of $F^{(-)}(x_{max},\theta,p)$ did not
depend on the choice of the rotation angle $\theta$ (indicative of the
accuracy of the method) provided the point $p$ is above the line
$(-\infty e^{-i\theta},+\infty e^{-i\theta})$.\\

Using Newton's method, we have located one zero of
$F^{(-)}(x_{max},\theta,p)$ for each of the states $^3n(1/2^-)$, $^4n(0^+)$,
$^4n(1^+)$, and $^4n(2^+)$ in the fourth quadrant of the complex $p$--plane,
for the Yukawa as well as Malfliet--Tjon $nn$--potential. The coordinates
$p_0$ of these zeros are given in Table 1. All zeros found lie below the
diagonal of the quadrant, which represents the threshold energy
$({\rm Re\,}E=0)$; that is the corresponding energies, $E_0=p_0^2/2m$, have negative real
parts. This implies that these zeros are sub--threshold resonances.\\

It is evident from the results of Table 1 that the position
of zeros are sensitive to the choice of the $nn$--potential.
The movement of these zeros with an artificial increase of the
potential strength also exhibits an extreme sensitivity to the choice of
the potential. Thus, to generate $^3n$ and $^4n$ bound states,
the Yukawa potential needs to be multiplied by 2.67 and 2.32 respectively,
while for the Malfliet--Tjon  potential by 4.04 and 3.59.\\

In Fig. 1 we show the movement of the three--neutron Jost function
zeros on the momentum plane, when the $nn$--potential is
multiplied by a  factor $\alpha$, $V_{nn}\longrightarrow\alpha V_{nn}$.
Open and filled circles correspond to the Yukawa and the Malfliet--Tjon
potentials  respectively. The zeros furthest from the origin correspond to
the physical potentials $(\alpha=1)$ and represent the zeros given in the
Table 1. The upward sequences of points are shown for
the uniform increase of  $\alpha$,  $\alpha=1.0, 1.1, 1.2,\cdots$.
In Fig. 2 the corresponding trajectory in the energy plane is depicted.
In Figs. 3 and 4 the results for the four--neutron
system are plotted  the notation being the same as for Figs. 1 and 2. It is
worth mentioning that a trajectory similar to these figures, was obtained
in Ref. \cite{yazaki} for a sub--threshold resonance
of the hypernuclear system $\Lambda nn$.\\

There have been several previous theoretical attempts to locate three
and four neutron resonances: using the  hyperspherical approach
\cite{gorb,birger,badalyan4,gutich,gorb1}, by an analytical continuation
of the Faddeev kernel \cite{glockle}, and by using the complex scaling
method \cite{csoto}. Invariably the authors of these papers searched for
multineutron resonances above the threshold and close to the real energy
axis. Only one of these attempts turned out to be successful, the
complex--scaling calculation by Csoto, Oberhummer, and Pichler
\cite{csoto} where they found a three--neutron resonance at
$E=(14-i6.5)$\,MeV for the $J^\pi=\frac32^+$  state and that all other
states up to $J=\frac52$ are nonresonant.\\

In short, all the above authors agree that there are no resonances in
the four neutron system, nor in the $\frac12^-$ state of three--neutrons. On
the other hand, it has been shown many times that by an artificial increase
of the neutron--neutron attraction, one can always obtain bound states for
the $^3n(\frac12^-)$ and $^4n(0^+)$ systems. Since the corresponding poles
cannot vannish or undergo a discontinuous relocation, they must be smoothly
move somewhere after decreasing the potential strength to its physical value.
Thus, to say that the states $^3n(\frac12^-)$ and $^4n(0^+)$ are nonresonant is
unsatisfactory.  One would like to know in which direction and how far the
poles move. Our present work sheds some light on this problem.\\

Using the minimal approximation $L=L_{minn}$ means that the actual position
of resonances in the complex momentum plane may be different from those
shown in the figures. In this connection, however,  we emphasize that in
bound states calculations the minimal approximation always underbinds the
system and that the omitted higher harmonics amount to an additional
attraction in the effective potential \cite{gorb}. Since in this
work we found that an increase in the attraction moves the resonances,
in the energy plane, up (decreases their width) and to the
right (closer to the threshold), the values given in Table 1 can
be considered as lower bounds for the
energies and upper bound for the widths of the resonances.

Since the physical potentials generate negative real parts of $E_0$,
the corresponding resonances can be excited at negative energies only,
that is when the $^3n$ or the $^4n$ system is placed in a sufficiently
strong attractive external field.  Such a situation can be realized,
for example, inside a nucleus. Besides the negative energy, the interior
of a nucleus provides a high frequency of  multineutron collisions due to
the high density of nucleons. Therefore, three or four neutrons inside a
nucleus could form an unstable cluster corresponding to a sub--threshold
resonance.\\

In conclusion, our approach which is based on the Jost function
calculation, is a power method which enables us to investigate
the analytical properties of the $S$--matrix in the complex
momentum plane. This opens up new possibilities in locating sub--threshold
resonances which is a difficult task for many other methods.
 Moreover the formalism  given  can be extended to
complex values of the angular momentum $\lambda$ and therefore
Regge trajectories can also be located.
\section*{Acknowledgments}
Financial support from the  University of South
Africa  and the Joint Institute  for Nuclear Research, Dubna,
is greatly appreciated.

\newpage
\begin{table}
\begin{tabular}{r|c|c|c|c|}
potential & \multicolumn{2}{c|}{Yukawa} &
\multicolumn{2}{c|}{Malfliet--Tjon}\\
\hline
 & $p_0\ ({\rm fm}^{-1})$ & $E_0\ ({\rm MeV})$ &
   $p_0\ ({\rm fm}^{-1})$ & $E_0\ ({\rm MeV})$ \\
\hline
$nn\ (0^+)$ & -i0.0401 & -0.0667 & -i0.0401 & -0.0667\\
$nnn\ (\frac12^-)$ & $0.232 -i0.560$ & $-5.39-i5.39$ & $0.291-i0.569$ &
	 $-4.95-i6.87$\\
$nnnn\ (0^+)$ & $0.337 -i0.594$ & $-4.96-i8.30$ & $0.402-i0.537$ &
	 $-2.64-i8.95$\\
$nnnn\ (1^+)$ & $0.262 -i0.790$ & $-11.5-i8.59$ & $0.435-i0.783$ &
	 $-8.77-i14.1$\\
$nnnn\ (2^+)$ & $0.467 -i0.744$ & $-6.97-i14.4$ & $0.535-i0.699$ &
	 $-4.19-i15.5$\\
\end{tabular}
\vspace{0.5cm}
\caption{Jost function zeros in  momentum and energy planes.}
\label{tab1}
\end{table}

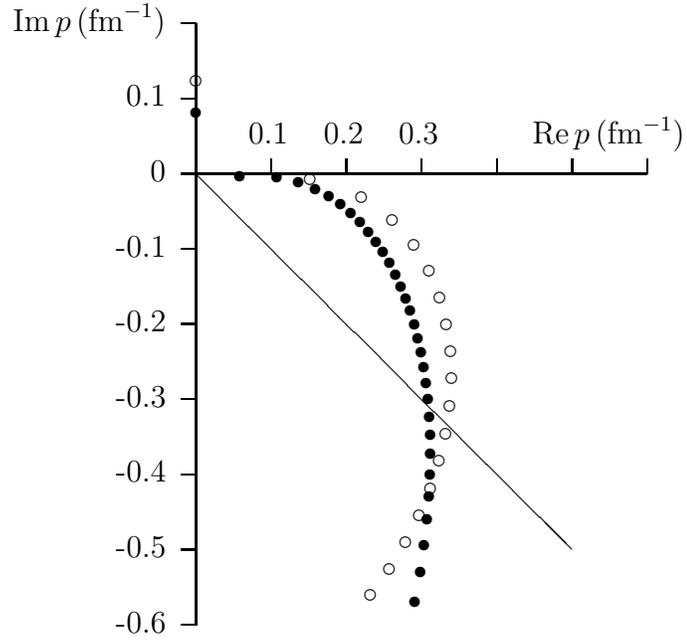
\begin{figure}
\begin{center}
\unitlength=1.00mm
\begin{picture}(70,100)
\put(20,70){%
\begin{picture}(0,0)%
\put(0,0){\line(1,0){60}}
\put(0,-60){\line(0,1){80}}
\multiput(10,0)(10,0){6}{\line(0,1){2}}
\multiput(-2,-60)(0,10){9}{\line(1,0){2}}
\put(0,0){\line(1,-1){50}}
\put(65,4){\llap{${\rm Re}\,p\,({\rm fm}^{-1})$}}
\put(-4,19){\llap{${\rm Im}\,p\,({\rm fm}^{-1})$}}
\put(-4,9){\llap{0.1}}
\put(-4,-1){\llap{0}}
\put(-4,-11){\llap{-0.1}}
\put(-4,-21){\llap{-0.2}}
\put(-4,-31){\llap{-0.3}}
\put(-4,-41){\llap{-0.4}}
\put(-4,-51){\llap{-0.5}}
\put(-4,-61){\llap{-0.6}}
\put(7,4){0.1}
\put(17,4){0.2}
\put(27,4){0.3}
\put(23.20,-56.03){\circle{1.30}}
\put(25.68,-52.54){\circle{1.30}}
\put(27.83,-49.00){\circle{1.30}}
\put(29.64,-45.42){\circle{1.30}}
\put(31.14,-41.81){\circle{1.30}}
\put(32.32,-38.18){\circle{1.30}}
\put(33.19,-34.53){\circle{1.30}}
\put(33.74,-30.89){\circle{1.30}}
\put(33.96,-27.24){\circle{1.30}}
\put(33.83,-23.61){\circle{1.30}}
\put(33.32,-19.99){\circle{1.30}}
\put(32.39,-16.41){\circle{1.30}}
\put(30.97,-12.89){\circle{1.30}}
\put(28.95,-9.45){\circle{1.30}}
\put(26.11,-6.15){\circle{1.30}}
\put(22.01,-3.12){\circle{1.30}}
\put(15.16,-0.67){\circle{1.30}}
\put(0.00,12.33){\circle{1.30}}
\put(29.13,-56.88){\circle*{1.30}}
\put(29.82,-52.96){\circle*{1.30}}
\put(30.33,-49.36){\circle*{1.30}}
\put(30.71,-46.03){\circle*{1.30}}
\put(30.96,-42.93){\circle*{1.30}}
\put(31.11,-40.04){\circle*{1.30}}
\put(31.16,-37.32){\circle*{1.30}}
\put(31.13,-34.75){\circle*{1.30}}
\put(31.02,-32.32){\circle*{1.30}}
\put(30.84,-30.02){\circle*{1.30}}
\put(30.60,-27.83){\circle*{1.30}}
\put(30.29,-25.74){\circle*{1.30}}
\put(29.92,-23.75){\circle*{1.30}}
\put(29.50,-21.84){\circle*{1.30}}
\put(29.02,-20.01){\circle*{1.30}}
\put(28.48,-18.25){\circle*{1.30}}
\put(27.89,-16.56){\circle*{1.30}}
\put(27.23,-14.94){\circle*{1.30}}
\put(26.52,-13.37){\circle*{1.30}}
\put(25.73,-11.87){\circle*{1.30}}
\put(24.88,-10.42){\circle*{1.30}}
\put(23.96,-9.04){\circle*{1.30}}
\put(22.94,-7.70){\circle*{1.30}}
\put(21.83,-6.43){\circle*{1.30}}
\put(20.60,-5.22){\circle*{1.30}}
\put(19.22,-4.07){\circle*{1.30}}
\put(17.66,-2.99){\circle*{1.30}}
\put(15.84,-2.00){\circle*{1.30}}
\put(13.63,-1.13){\circle*{1.30}}
\put(10.72,-0.43){\circle*{1.30}}
\put(5.82,-0.30){\circle*{1.30}}
\put(0.00,8.0812345659360){\circle*{1.30}}
\end{picture}%
}
\end{picture}
\end{center}
\caption{Movement of the $3n$ Jost function zeros when the $nn$--potential
is multiplied by the enhancing factor
$V_{nn}\longrightarrow\alpha V_{nn}$. Open and filled circles correspond to
Yukawa and Malfliet--Tjon potentials respectively. The sequence of points are
shown for a uniform increase of $\alpha$ by 0.1, $\alpha=1.0
{\rm (lowest point)}, 1.1, 1.2, ...$.}
\label{p3n}
\end{figure}
\newpage
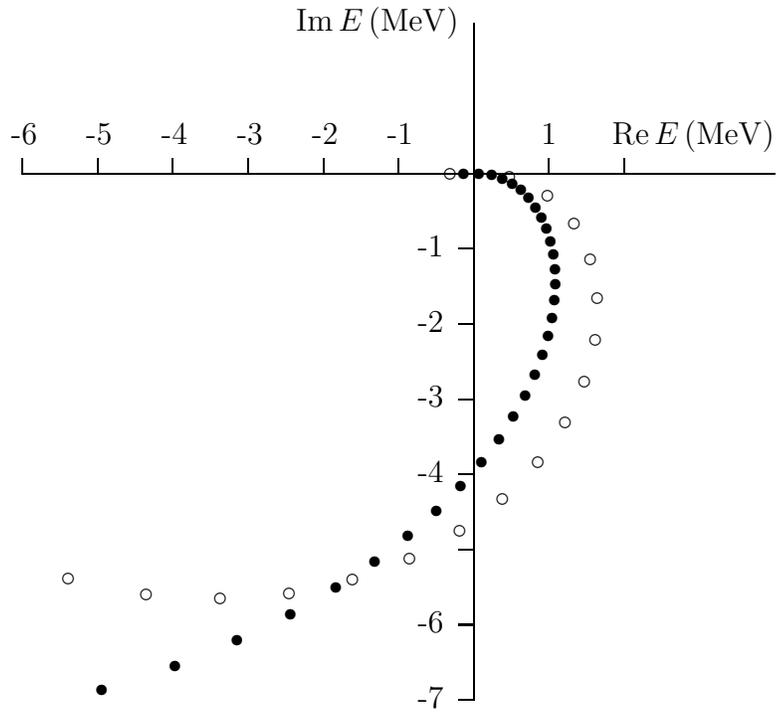
\begin{figure}
\begin{center}
\unitlength=1.00mm
\begin{picture}(120,110)
\put(60,85){%
\begin{picture}(0,0)%
\put(-60,0){\line(1,0){100}}
\put(0,-70){\line(0,1){90}}
\multiput(-60,0)(10,0){9}{\line(0,1){2}}
\multiput(-2,-70)(0,10){7}{\line(1,0){2}}
\put(40,4){\llap{${\rm Re}\,E\,({\rm MeV})$}}
\put(-2,19){\llap{${\rm Im}\,E\,({\rm MeV})$}}
\put(-4,-11){\llap{-1}}
\put(-4,-21){\llap{-2}}
\put(-4,-31){\llap{-3}}
\put(-4,-41){\llap{-4}}
\put(-4,-61){\llap{-6}}
\put(-4,-71){\llap{-7}}
\put(-61.50,4){-6}
\put(-51.50,4){-5}
\put(-41.50,4){-4}
\put(-31.50,4){-3}
\put(-21.50,4){-2}
\put(-11.50,4){-1}
\put(9,4){1}
\put(-53.90,-53.86){\circle{1.30}}
\put(-43.54,-55.93){\circle{1.30}}
\put(-33.71,-56.51){\circle{1.30}}
\put(-24.54,-55.80){\circle{1.30}}
\put(-16.12,-53.96){\circle{1.30}}
\put(-8.55,-51.14){\circle{1.30}}
\put(-1.89,-47.50){\circle{1.30}}
\put(3.82,-43.18){\circle{1.30}}
\put(8.52,-38.33){\circle{1.30}}
\put(12.16,-33.09){\circle{1.30}}
\put(14.72,-27.61){\circle{1.30}}
\put(16.15,-22.03){\circle{1.30}}
\put(16.43,-16.54){\circle{1.30}}
\put(15.51,-11.34){\circle{1.30}}
\put(13.34,-6.66){\circle{1.30}}
\put(9.84,-2.84){\circle{1.30}}
\put(4.75,-0.42){\circle{1.30}}
\put(-3.15,0.){\circle{1.30}}
\put(-49.46,-68.673720581702){\circle*{1.30}}
\put(-39.703568334183,-65.442439068809){\circle*{1.30}}
\put(-31.426152760078,-62.048380241240){\circle*{1.30}}
\put(-24.370627642128,-58.575979506996){\circle*{1.30}}
\put(-18.336958103482,-55.083100189318){\circle*{1.30}}
\put(-13.160809103253,-51.614774183350){\circle*{1.30}}
\put(-8.7406934560743,-48.187002466198){\circle*{1.30}}
\put(-4.9493565737669,-44.827523754598){\circle*{1.30}}
\put(-1.7141748046447,-41.553510676769){\circle*{1.30}}
\put(1.0322036855025,-38.372384560925){\circle*{1.30}}
\put(3.3476330733283,-35.291226036414){\circle*{1.30}}
\put(5.2802008997774,-32.315778515726){\circle*{1.30}}
\put(6.8704128727660,-29.450137295439){\circle*{1.30}}
\put(8.1528946021925,-26.697267683085){\circle*{1.30}}
\put(9.1574520330527,-24.059476292410){\circle*{1.30}}
\put(9.9098420849812,-21.538686224969){\circle*{1.30}}
\put(10.436442855380,-19.140945388089){\circle*{1.30}}
\put(10.743628024444,-16.857664477867){\circle*{1.30}}
\put(10.861964571331,-14.695569754639){\circle*{1.30}}
\put(10.803390540723,-12.659289860255){\circle*{1.30}}
\put(10.579017765652,-10.750704075471){\circle*{1.30}}
\put(10.199836394862,-8.9717482722626){\circle*{1.30}}
\put(9.6759807874705,-7.3256152949260){\circle*{1.30}}
\put(9.0160717675616,-5.8170087114400){\circle*{1.30}}
\put(8.2268805485975,-4.4521913830301){\circle*{1.30}}
\put(7.3131135732625,-3.2392475031393){\circle*{1.30}}
\put(6.2769810128442,-2.1887771148406){\circle*{1.30}}
\put(5.1170357464858,-1.3152272812095){\circle*{1.30}}
\put(3.8250921246682,-0.63921417265143){\circle*{1.30}}
\put(2.3771695331332,-0.19108523278243){\circle*{1.30}}
\put(0.70086600181794,-0.0072155282598370){\circle*{1.30}}
\put(-1.3532298472602,0.00){\circle*{1.30}}
\end{picture}%
}
\end{picture}
\end{center}
\caption{Movement of the three--neutron Jost function zeros on the energy
plane. The notation is the same as in Fig. 1. }
\label{e3n}
\end{figure}
\newpage
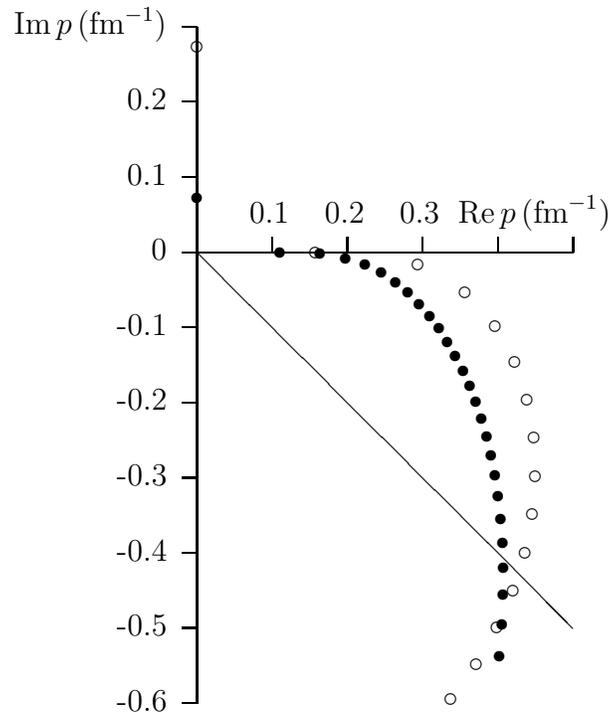
\begin{figure}
\begin{center}
\unitlength=1.00mm
\begin{picture}(90,130)
\put(30,75){%
\begin{picture}(0,0)%
\put(0,0){\line(1,0){50}}
\put(0,-60){\line(0,1){90}}
\multiput(10,0)(10,0){5}{\line(0,1){2}}
\multiput(-2,-60)(0,10){10}{\line(1,0){2}}
\put(0,0){\line(1,-1){50}}
\put(55,4){\llap{${\rm Re}\,p\,({\rm fm}^{-1})$}}
\put(-4,29){\llap{${\rm Im}\,p\,({\rm fm}^{-1})$}}
\put(-4,19){\llap{0.2}}
\put(-4,9){\llap{0.1}}
\put(-4,-1){\llap{0}}
\put(-4,-11){\llap{-0.1}}
\put(-4,-21){\llap{-0.2}}
\put(-4,-31){\llap{-0.3}}
\put(-4,-41){\llap{-0.4}}
\put(-4,-51){\llap{-0.5}}
\put(-4,-61){\llap{-0.6}}
\put(7,4){0.1}
\put(17,4){0.2}
\put(27,4){0.3}
\put(33.712707597449,-59.428869855879){\circle{1.30}}
\put(37.094056320806,-54.759351177074){\circle{1.30}}
\put(39.844982727576,-49.926487631386){\circle{1.30}}
\put(42.000487304039,-44.976824492793){\circle{1.30}}
\put(43.578232055910,-39.945503210943){\circle{1.30}}
\put(44.580635386010,-34.861634728165){\circle{1.30}}
\put(44.994310323856,-29.752263806337){\circle{1.30}}
\put(44.786794601583,-24.646051595785){\circle{1.30}}
\put(43.899020329787,-19.578104980857){\circle{1.30}}
\put(42.229206240314,-14.598447933255){\circle{1.30}}
\put(39.595580818605,-9.7905945087857){\circle{1.30}}
\put(35.632714243742,-5.3219430482462){\circle{1.30}}
\put(29.382960787811,-1.6218360493069){\circle{1.30}}
\put(15.734662280249,-0.0085290371484){\circle{1.30}}
\put(0.00,27.397){\circle{1.30}}
\put(40.202153637033,-53.744080036587){\circle*{1.30}}
\put(40.553870856081,-49.502745048875){\circle*{1.30}}
\put(40.719926290217,-45.597499152254){\circle*{1.30}}
\put(40.731288015832,-41.979263671937){\circle*{1.30}}
\put(40.610782794652,-38.609057894967){\circle*{1.30}}
\put(40.375446657460,-35.455498007900){\circle*{1.30}}
\put(40.038067591547,-32.493028777193){\circle*{1.30}}
\put(39.608223281562449092,-29.700645059856017882){\circle*{1.30}}
\put(39.092978126049665599,-27.060959703873477267){\circle*{1.30}}
\put(38.497355484739148590,-24.559513370231583762){\circle*{1.30}}
\put(37.824653516994677371,-22.184273589089142931){\circle*{1.30}}
\put(37.076640620187401032,-19.925252366022366823){\circle*{1.30}}
\put(36.253640958690158680,-17.774240591245557552){\circle*{1.30}}
\put(35.354513329517328613,-15.724618685298222265){\circle*{1.30}}
\put(34.376552298240148353,-13.771295741526703993){\circle*{1.30}}
\put(33.315199129493233698,-11.910685786759636628){\circle*{1.30}}
\put(32.163689588479632553,-10.140887580873264584){\circle*{1.30}}
\put(30.912325769188525593,-8.4617143173957756752){\circle*{1.30}}
\put(29.547172467281396235,-6.8754774056484169176){\circle*{1.30}}
\put(28.048158624877439493,-5.3876403768336064704){\circle*{1.30}}
\put(26.385324175812013836,-4.0084277260368415818){\circle*{1.30}}
\put(24.511395959359613683,-2.7568197846802467571){\circle*{1.30}}
\put(22.345941981343278582,-1.6626901466069744784){\circle*{1.30}}
\put(19.732367279478,-0.78091732259787){\circle*{1.30}}
\put(16.310721916034,-0.20376537574848){\circle*{1.30}}
\put(10.982585298359,-0.033729576615428){\circle*{1.30}}
\put(0.,7.2187750339508){\circle*{1.30}}
\end{picture}%
}
\end{picture}
\end{center}
\caption{Movement of the four--neutron Jost function zeros on the momentum
plane. The notation is the same as in Fig. 1.}
\label{p4n}
\end{figure}
\newpage
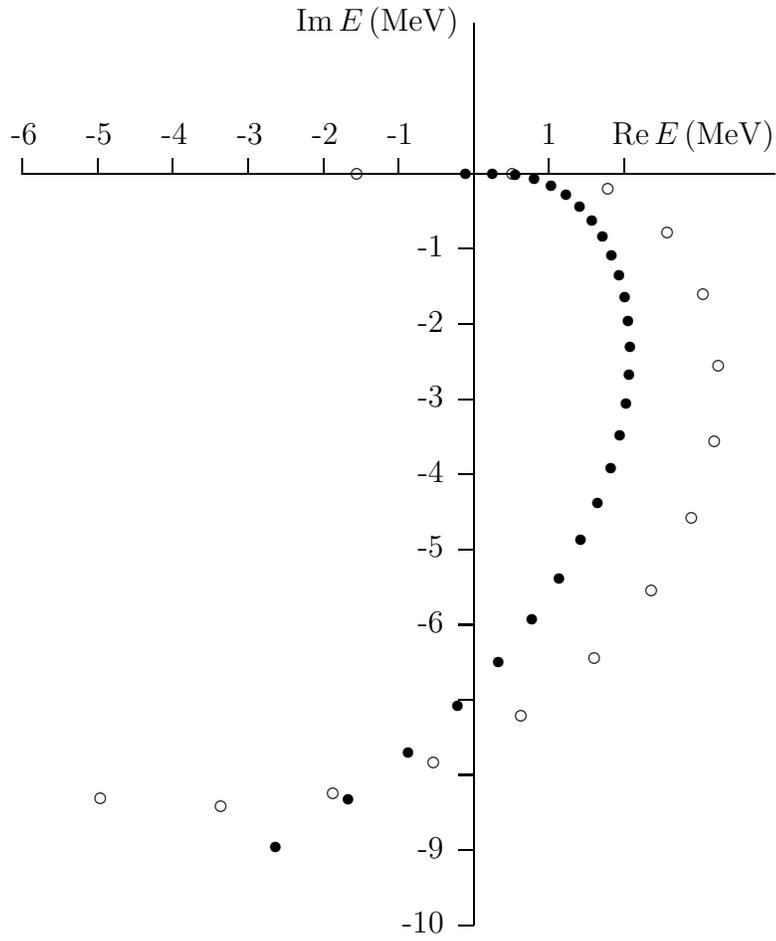
\begin{figure}
\begin{center}
\unitlength=1.00mm
\begin{picture}(120,140)
\put(60,115){%
\begin{picture}(0,0)%
\put(-60,0){\line(1,0){100}}
\put(0,-100){\line(0,1){120}}
\multiput(-60,0)(10,0){9}{\line(0,1){2}}
\multiput(-2,-100)(0,10){10}{\line(1,0){2}}
\put(40,4){\llap{${\rm Re}\,E\,({\rm MeV})$}}
\put(-2,19){\llap{${\rm Im}\,E\,({\rm MeV})$}}
\put(-4,-11){\llap{-1}}
\put(-4,-21){\llap{-2}}
\put(-4,-31){\llap{-3}}
\put(-4,-41){\llap{-4}}
\put(-4,-51){\llap{-5}}
\put(-4,-61){\llap{-6}}
\put(-4,-91){\llap{-9}}
\put(-4,-101){\llap{-10}}
\put(-61.50,4){-6}
\put(-51.50,4){-5}
\put(-41.50,4){-4}
\put(-31.50,4){-3}
\put(-21.50,4){-2}
\put(-11.50,4){-1}
\put(9,4){1}
\put(-49.632470007504,-83.030421672806){\circle{1.30}}
\put(-33.622678304559,-84.179968517249){\circle{1.30}}
\put(-18.753392651397,-82.442432115473){\circle{1.30}}
\put(-5.3641912602386,-78.286929001735){\circle{1.30}}
\put(6.2872254411892,-72.141246486964){\circle{1.30}}
\put(15.998873333887,-64.408048457388){\circle{1.30}}
\put(23.607539101847,-55.478377799553){\circle{1.30}}
\put(28.977212043263,-45.744983218856){\circle{1.30}}
\put(31.989931443323,-35.618171413015){\circle{1.30}}
\put(32.536338332004,-25.548519696883){\circle{1.30}}
\put(30.500746106800,-16.065783870491){\circle{1.30}}
\put(25.722692436487,-7.8589633962615){\circle{1.30}}
\put(17.835368632734,-1.9749160679767){\circle{1.30}}
\put(5.1301594878526,-0.0055616489157397){\circle{1.30}}
\put(-15.554,0.00){\circle{1.30}}
\put(-26.361854912969060472,-89.541855665441385526){\circle*{1.30}}
\put(-16.699373122639964429,-83.197013030567124048){\circle*{1.30}}
\put(-8.7240548336597756318,-76.947434633463629439){\circle*{1.30}}
\put(-2.1388640158743671416,-70.861297242303740163){\circle*{1.30}}
\put(3.2859011830067003102,-64.979550619584518500){\circle*{1.30}}
\put(7.7307775859482030789,-59.326273272124421254){\circle*{1.30}}
\put(11.339707780482326527,-53.914979180867792508){\circle*{1.30}}
\put(14.228937123917044794,-48.752555070697782824){\circle*{1.30}}
\put(16.493405313107130095,-43.841771006626446550){\circle*{1.30}}
\put(18.211431509761726311,-39.182919589492728463){\circle*{1.30}}
\put(19.448195127809579930,-34.774933590677035156){\circle*{1.30}}
\put(20.258379345399140092,-30.616133742991711308){\circle*{1.30}}
\put(20.688164293395470139,-26.704768782515926340){\circle*{1.30}}
\put(20.776750283124263419,-23.039397855536694060){\circle*{1.30}}
\put(20.557535555723469578,-19.619288856092398365){\circle*{1.30}}
\put(20.058967464417429838,-16.444675932746248836){\circle*{1.30}}
\put(19.305278697213261729,-13.517238251264973581){\circle*{1.30}}
\put(18.316997890322674891,-10.840172133325096659){\circle*{1.30}}
\put(17.110853510297383551,-8.4190855895443317269){\circle*{1.30}}
\put(15.699928927986799287,-6.2625195142038947438){\circle*{1.30}}
\put(14.092898006014207279,-4.3831126251054552956){\circle*{1.30}}
\put(12.292026582173731786,-2.8004160676826139742){\circle*{1.30}}
\put(10.289692103469543216,-1.5397709727005329250){\circle*{1.30}}
\put(8.0555239110988,-0.63860216312463){\circle*{1.30}}
\put(5.5118161706632,-0.13773671197454){\circle*{1.30}}
\put(2.4993162690137,-0.015351881627365){\circle*{1.30}}
\put(-1.0797995892148,0.){\circle*{1.30}}
\end{picture}%
}
\end{picture}
\end{center}
\caption{Movement of the four--neutron Jost function zeros on the energy
plane. The notation is the same as in Fig. 1.}
\label{e4n}
\end{figure}

\end{document}